\documentclass[11pt]{article}

\usepackage{floatflt}
\usepackage{jvink}
\usepackage{times}
\usepackage{psfig}
\usepackage{graphicx}

\usepackage[sectionbib]{natbib}
\usepackage{natbibspacing}
\newcommand{\adspr}{{\it Adv. Space Res.}}
\newcommand{\apj}{{\it Astrophys. J.}}

\newcommand{\apjl}{{\it Astrophys. J.}}
\newcommand{\aap}{{\it Astron. Astroph.}}
\newcommand{\nat}{{\it Nature}}
\newcommand{\jetp}{{\it J. Theoretical and Experimental Phys.}}
\newcommand{\mnras}{{\it Mon. Not. Roy. Astron. Soc.}}

\newcommand{\araa}{{\it Ann. Rev. of Astron  Astroph.}}
\newcommand{\pasj}{{\it Publ. Astron. Soc. Japan}}

\newcommand{\xmm}{{\it XMM-Newton}}
\newcommand{\chandra}{{\it Chandra}}
\newcommand{\asca}{{\it ASCA}}
\newcommand{\rosat}{{\it ROSAT}}
\newcommand{\cangeroo}{{\it CANGAROO}}
\newcommand{\whipple}{{\it Whipple}}
\newcommand{\hegra}{{\it HEGRA}}
\newcommand{\hess}{{\it HESS}}
\newcommand{\sax}{{\it BeppoSAX}}
\newcommand{\rxte}{{\it RXTE}}

\newcommand{\egret}{{\it CGRO-EGRET}}
\newcommand{\integral}{{\it INTEGRAL}}
\newcommand{\glast}{{\it GLAST}}

\newcommand{\rcwes}{{RCW~86}}
\newcommand{\casa}{{Cas~A}}
\newcommand{\snts}{{SN~1006}}
\newcommand{\rxj}{{RXJ 1713.7-3946}}

\newcommand{\gray}{{$\gamma$-ray}}

\newcommand{\Halpha}{{H$\alpha$}}

\newcommand{\Te}{$T_{\rm e}$}
\newcommand\arcmin{\mbox{$^\prime$}}%
\newcommand\arcsec{\mbox{$^{\prime\prime}$}}%

\title{SHOCKS AND PARTICLE ACCELERATION IN SUPERNOVA~REMNANTS: OBSERVATIONAL FEATURES}
\author{Jacco Vink \footnote{Columbia Astrophysics Laboratory, 
MC 5247, 550 W 120$^{th}$\ Street, 
New York, NY 10027 , USA (jvink@astro.columbia.edu)}
\footnote{Chandra fellow}
}

\begin{document}

\maketitle

\begin{abstract}
The last ten years a number of observational advances have substantially
increased our knowledge of shock phenomena in supernova remnants. 
This progress has mainly been made possible by the recent improvements
in X-ray and \gray\ instrumentation.
It has become clear that some shell-type supernova remnants, e.g. \snts, 
have X-ray emission
dominated by synchrotron radiation, proving that electrons are accelerated
up to 100~TeV. 
This is still an order of magnitude below  $3\times10^{15}$~eV,
at which energy the ion cosmic ray spectrum at earth shows a 
spectral break. So one of the major goals is to prove that supernova remnants
are capable of accelerating ions at least up that energy.

Here I review the evidence that ions and electrons are accelerated 
up to energies $\sim 100$~TeV in supernova remnants, 
and, in addition, the recent progress that has been  made in
understanding the physics of collisionless shock fronts and the magnetic fields
inside supernova remnants.
\end{abstract}

\section{Introduction}
Supernova remnant shocks are considered to be the prime source of cosmic rays 
for energies at least up to the ``knee'' at $\sim 3\times10^{15}$~eV 
and probably up to the ``ankle'' at $\sim10^{18}$~eV 
(above which the cosmic rays are thought to be of  extra-galactic origin).
One of the main reasons is that supernova remnants (SNRs) are the only galactic
sources that are able to provide the energy necessary to maintain the 
observed cosmic ray density, $10^{48}$~erg/yr, 
if about 10\% of the average supernova kinetic energy is dumped into cosmic ray
acceleration, and assuming a supernova rate of 1 every 50 - 100~yrs 
\citep[e.g.][]{longairv2}.
Another reason is that radio synchrotron emission from SNRs have spectral 
indexes implying a spectral energy slope of $\sim -2.2$, 
consistent with the locally observed cosmic ray spectral index,
which has a slightly steeper index due to propagation effects.

For a long time the observational evidence that SNRs are indeed the source of 
cosmic rays consisted solely of the observation of radio synchrotron emission,
caused by the presence of relativistic electrons.
In some cases SNRs may just light up the relativistic galactic background
electrons by means of compressing the local magnetic field \citep{vdlaan62}, 
but for the very bright, young, SNRs,
the most likely explanation is that the electrons are accelerated by SNR shocks.
A good example is the youngest and brightest known galactic SNR, 
Cassiopeia A (\casa).

Although for these reasons it is likely that SNRs accelerate electrons to 
relativistic energies, at least up to several GeV, 
radio synchrotron emission does not prove that SNRs are capable of accelerating 
cosmic rays up to the ``knee'', and, moreover, it only proves that 
{\em electrons} are being accelerated, 
whereas the cosmic rays observed at earth are dominated by ions.
However, over the last decade  evidence has been accumulating that cosmic rays 
are indeed being accelerated up to the ``knee''.
This progress has been made possible by the coming of age of 
X-ray astronomy, with missions like \asca, \sax, \rxte, and, 
more recently \chandra\ and \xmm.
In addition, a lot has been learned from \gray\ observations with space
based instruments like \egret\ and ground based Cherenkov telescopes
such as \whipple\, \hegra, and \cangeroo.

Although the detection of ion cosmic rays in SNRs can be considered one of 
the main goals of cosmic ray physics,
another important, but still poorly understood, aspect of cosmic ray physics 
is the injection mechanism at low energies.
This process may be intimately connected to shock heating and temperature equilibration by collisionless shocks.

\section{Shock structure and electron-ion tempererature equilibration}
The particle collision length scales in supernova remnant shocks
are much larger than the typical size of the shock structure.
For instance, the young supernova remnant \casa\ has a shock velocity of 
about 5000~km/s \citep{vink98a,delaney03} and a typical ion density 
of 5~cm$^{-3}$, implying that a proton entering the shock
has a relative energy of $\sim 8$~keV.
The deflection time scale for a proton, assuming proton proton interactions, is
$\tau_{\rm pp} \sim 800$~yr \citep[e.g.][]{nrlplasma}, 
more than the age of \casa, which is about $\sim 320$~yr!
So other, collective rather than two body, interactions, have to be at work
to heat the gas behind a supernova remnant shocks.
They are therefore referred to as {\em collisionless shocks}.

Although SNRs have collisionless shocks, the Hugoniot relations,
which are based on the conservation of mass, momentum, 
and energy, are still valid.
If we neglect, for the moment, the energy deposition in cosmic rays,
and radiative losses, the Hugoniot relations in the limit of high Mach number
shocks state
\citep[e.g.][]{zeldovich66,mckee80}:
\begin{eqnarray}
	\frac{\rho_{2,i}}{\rho_{1,i}} = \chi = \frac{\gamma +1}{\gamma -1} = 4 \\
	kT_{2,i} = \frac{2(\gamma-1)}{\gamma+1)^2} m_i v_s^2  = \frac{3}{16} m_i v_s^2,
	\label{eq-shocks}
\end{eqnarray}
where subscript $1$ and $2$ refer to the pre-shock and post-shock conditions for
particle species $i$; with $v_s$ the shock velocity. The numerical values were obtained for
$\gamma= 5/3$, the specific heat ratio for a monatomic gas. Note, however, that
for a relativistic gas (i.e. the pressure is dominated by relativistic cosmic rays) 
$\gamma = 4/3$, giving a compression ratio of $\chi = 7$.
The post-shock gas moves away from the shock front with a relative velocity of 
$u_2 = v_s/\chi$.

The shock jump conditions hold for each particle species $i$, and raises the question whether
the, still poorly understood, mechanism for shock heating the particles will also tend to rapidly equilibrate
the temperatures of the various particles involved.
In addition, if a large neutral fraction exists in the pre-shock material, or the shock is moving
through a medium with a high metal content (e.g. ejecta),
a large fraction of the electrons behind the shock may originate from ionizations. These so-called
secondary electrons have to be heated by primary electrons, resulting in an overall lower electron temperature
than indicated by Eq.~(\ref{eq-shocks}), see \citet{itoh84}.

The earliest indications that the electron and ion temperatures are indeed not equilibrated
are the relatively low electron temperatures inside SNRs,
which, according to X-ray observations, in no object seem to exceed 5~keV,
whereas a typical shock velocity of $4000$~km/s should give rise to
a mean plasma temperature of 19~keV.

More direct evidence for only modest equilibration of electrons and ions behind SNR shocks
comes from optical and UV spectroscopy.
\Halpha\ emission from non-radiative shocks in a number of SNRs is 
characterized
spectroscopically by a narrow component and a broad component 
\citep{ghavamian01}. 
The narrow component is thought to come from neutral hydrogen briefly 
excited after being overtaken by the shock. 
The broad component is the result of charge 
exchange between neutral and ionized hydrogen. 
The ionized hydrogen is heated according to Eq.~(\ref{eq-shocks}), 
and after charge exchange, which leaves the atom in an excited state, 
the thermal motion of the heated hydrogen gives rise to
Doppler broadening.
The width of the broad line component can be used to determine the shock
velocity, if the equilibration fraction is known (see Eq.~\ref{eq-shocks}).
The equilibration fraction is determined from the ratio between the
narrow and broad component \citep{ghavamian01}.

Several SNRs with broadened \Halpha\ emission are known 
\citep[e.g.][]{smith91,ghavamian01}.
In particular the work by \citet{ghavamian01,ghavamian02} provides clear evidence that 
the youngest SNRs,
i.e. the ones with large shock velocities such as SN~1006 and Tycho, 
have a large ratio between the electron and proton temperatures 
(\Te/$T_{\rm p} < 0.07$).
Additional proof for low equilibration  behind the shock front of \snts\
comes from UV spectra obtained with the Hopkins Ultraviolet Telescope,
which show broad C~IV, N~V and O~VI lines, indicating temperature
non-equilibration of these elements \citep{laming96}.

\begin{figure}
	\centerline{
		\psfig{figure=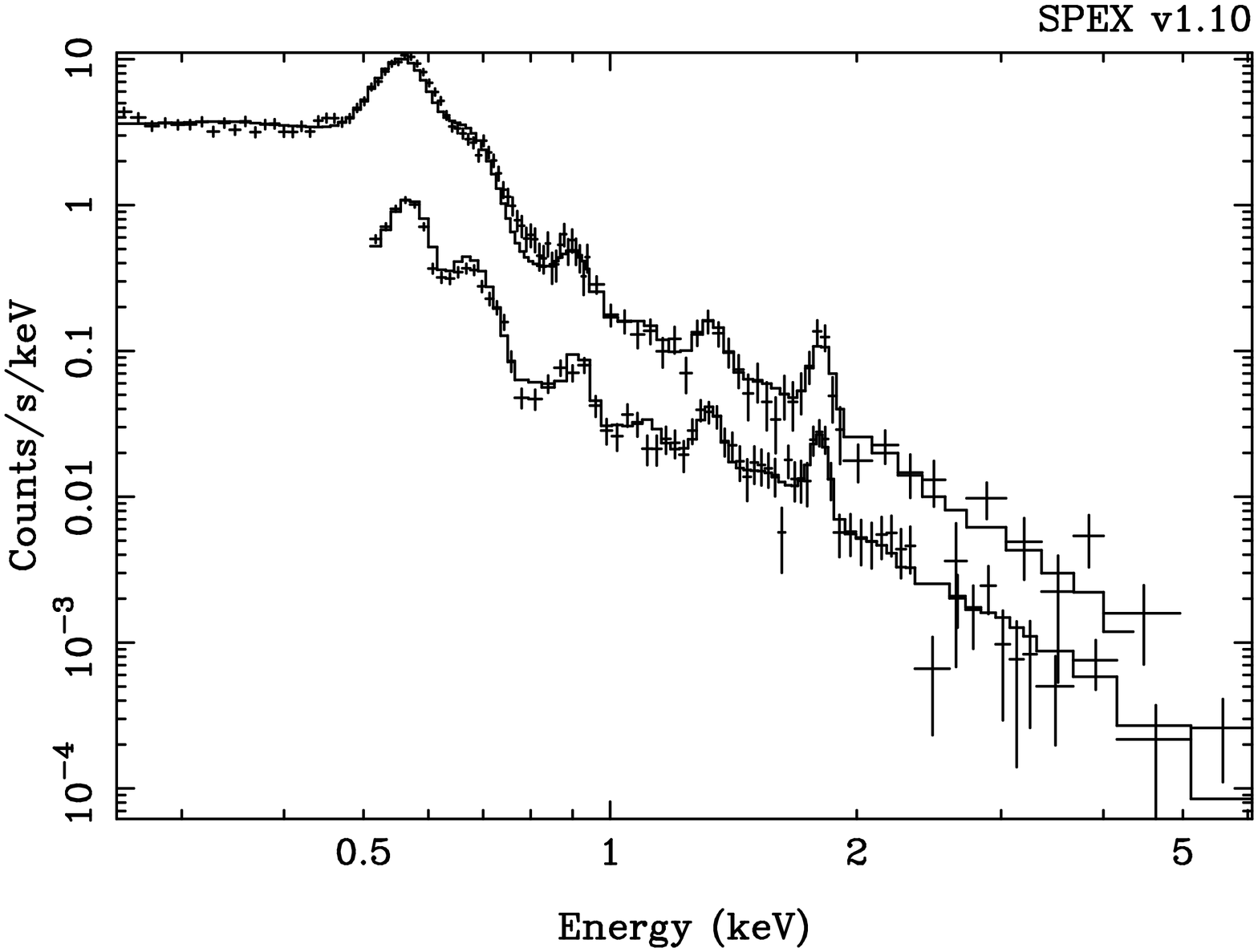,angle=-90,width=0.5\textwidth}
		\psfig{figure=sn1006_rgs_cospar.ps,angle=-90,width=0.5\textwidth}
	}
	\caption{
Evidence for non-equilibrated oxygen and electron 
temperatures behind the shock front in the northwest of
SN 1006 based on \xmm\ observations \citep{vink03b}. 
On the left \xmm\ EPIC-MOS (lower) and EPIC-PN 
(x5) spectra with best fit non-eqilibrium ionization model. 
On the right: \xmm\ RGS1 spectrum of the O~VII line emission.
The best fit model (solid line)
has a significant thermal broadening;
the best fit model without broadening is indicated by the dotted line.}
\end{figure}

Very recently \citet{vink03b} determined the amount of 
electron-oxygen equilibration from X-ray observations by \xmm.
The observation pointed at a bright, but narrow (0.4\arcmin\ FWHM) knot in
the northeast of SN~1006, which made it a good target for the \xmm\ reflective
grating spectrometer (RGS).
The size of the knot translates into a spectral broadening of 
$\Delta\lambda \sim 0.05$~\AA\ (FWHM) or a resolution power of 
$\lambda/\Delta\lambda \sim 430$ for the dominant
O~VII He$\alpha$ line emission. 
However, SN 1006 itself is very extended
(30\arcmin) and contributes to the spectrum as well.
What made it, nevertheless, a good target for measuring 
the thermal Doppler broadening
is that this knot is compact and lies at the edge of the remnant, making
it very likely that Doppler broadening is due to thermal motions rather
than internal bulk velocities.

Modeling of the O~VII line emission, which consists of three bright lines, 
revealed both a very low ionization 
parameter, $\log(n_{\rm e} t) \simeq 9.2$ (cgs units), and a substantial line
broadening with $\sigma_E = 3.4\pm0.5$~eV  at 540~eV (Fig.~\ref{fig-sn1006}). 
The detection of broadening
is statistically very significant ($>6.5\sigma$ level), and
corresponds to a temperature for the oxygen ions of
$kT = 528\pm150$~keV, or, according to Eq.~(\ref{eq-shocks}), 
a shock velocity of  $\sim 4100$~km/s.
This is comparable, but somewhat faster, than the shock velocity inferred
from the \Halpha\ emission ($\sim 3000$~km/s). This is not surprising
as the X-ray emission comes from further downstream the shock front,
where the plasma may have been shocked earlier, when the
shock velocity was higher than at present.
The electron temperature of the knot has been measured with
the CCD instruments of \xmm, and is \Te $=1.5\pm0.2$~keV.
Calculations of the plasma history show that the difference between
electron and oxygen temperature support only a small 
equilibration fraction near the shock front ($\sim$5\%), 
and subsequent equilibration proceeds through Coulomb interactions.
A conclusion that, at least for the case of \snts, is based on
optical, UV and X-ray spectroscopy.

\section{Cosmic ray acceleration and X-ray synchrotron emission from shell type SNRs}
The most likely mechanism for cosmic ray production in SNRs is
first order Fermi acceleration,
according to which an energetic particle repeatedly scatters upstream and 
downstream due to the presence of plasma waves. 
For each scattering the energy of the particle in the local frame is 
conserved, resulting in an increase in energy after each
crossing of the shock front, i.e. after each change of local frame 
\citep[see e.g.][]{bell78a,blandford78}.
As the flow in the post shock region has a tendency to sweep the 
particles away from the shock front with a probability of  $4u_2/v_s$, 
i.e. independent of the particle energy, the particle distribution
is an inverse power law in momentum.
The predicted power law index is 
$\Gamma = (2u_2 - u_1)/(u_1-u_2) = (\chi + 2)/(\chi-1)$, which corresponds
to $\Gamma = 2 $ for a shock compression ratio of $\chi = 4$.

The earliest evidence that SNRs are indeed sites of cosmic ray acceleration
was their non-thermal radio emission, which for the brightest remnants could
only be explained by assuming  recent electron cosmic ray acceleration.
The SNR radio emission is the result of synchrotron radiation caused by 
relativistic
electrons with energies in the MeV to GeV range.
Most shell-type SNRs have radio spectral indexes of $\alpha \sim 0.6$\
\citep{stephenson02}, but some remnant have steeper spectra,
like \casa, which has $\alpha = 0.77$,
steeper than predicted by simple first order Fermi acceleration
theory, $\alpha = (\Gamma-1)/2 = 0.5$.
This is probably the result of the back reaction of the cosmic ray pressure
on the shock structure. For a substantial cosmic ray pressure the flow ahead of
the shock will be slowed down, as seen from the shock frame. 
As a result a shock pre-cursor forms and the density jump is smoothed over 
larger length scales \citep[e.g.][]{eichler79,bell87}.
Moreover the density jump experienced by relativistic particles depends on the
particle's gyroradius, i.e. particle energy. Very energetic particles
will experience the full density jump from the undisturbed medium to the
fully shocked plasma, whereas less energetic particles will only experience
the difference between the shock precursor and the main shock, which may be
less than the canonical factor 4 (see Eq.~\ref{eq-shocks}).
As a result,
the radio spectral index should be steeper at low radio frequencies than
at high frequencies. 
This theory has been tested and confirmed for the Tycho and Kepler SNRs 
by \citet{reynolds92}. Very recently \citet{jones03} reported that
infrared observations of \casa\ also indicate a flattening of the
synchrotron spectrum at high frequencies.

An extreme case of cosmic ray modification of the shock structure
is one in which the pressure is completely dominated by cosmic rays. In that case
the shock structure is smooth instead of characterized by a 
distinct pressure jump. 
\citet{malkov97} has shown that this will result in a particle index  $\Gamma=1.5$.
It is not clear how physical this solution of the couple gas/cosmic ray model
is, but there is no observational evidence that some SNR shocks
are characterized by this extreme case of cosmic ray acceleration, which should
be distinguishable by a rather flat radio index of $\alpha=0.25$ and 
large energy losses due to escaping high energy cosmic rays 
(for $\Gamma\leq 2$ the highest energy cosmic rays carry most of the energy).
$\Gamma=1.5$ is also the spectral index expected for a relativistic
gas with $\chi = 7$, but \citet{malkov97} has shown that for 
cosmic ray dominated shocks $\Gamma=1.5$ even for  $\chi > 7$.

\begin{figure}
\parbox{0.5\textwidth}{
		\psfig{figure=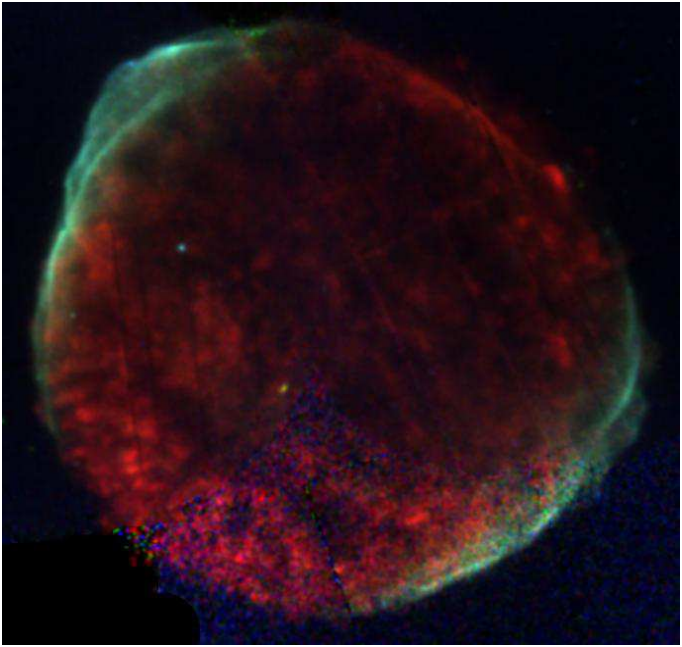,width=0.5\textwidth}}
\hskip 0.05\textwidth
\parbox{0.45\textwidth}{
	\vskip 5cm
	\caption{
\xmm\ (EPIC instruments) image of SN 1006, RGB color coded according
to energy:  0.50-0.61~keV (red),
0.75 - 1.6~keV (green), and 2.0-7 keV (blue). 
Regions that show up bluish are dominated by synchrotron emission, whereas
red regions have substantial O~VII line emission.
The image is based on mosaics of several observations;
the exposure of the southwestern part was shorter, and is therefore noisier.
\label{fig-sn1006}}}
\end{figure}

Although SNRs have the available energy to mark them as the primary sites
of cosmic ray acceleration, one long standing questions is if they
are capable to accelerate cosmic rays at least up to the observed break in the
cosmic ray spectrum at $3\times10^{15}$~eV \citep{lagage83}.
A partial answer has been provided by the recent discovery that some shell-type
SNRs, which usually are dominated by thermal X-ray emission, in fact also emit
X-ray synchrotron radiation.
This was first discovered in \snts.

For a long time the featureless X-ray spectrum of \snts\ was difficult to
explain by thermal radiation, as thermal radiation
produces line emission.
An interpretation invoking carbon rich, hot plasma, which suppresses line
emission above 0.5~keV \citep{hamilton86b}, became untenable after \asca\ observations 
showed that \snts\ does display line emission typical for hot plasma,
but that the overall X-ray emission is dominated by continuum emission coming 
from the northeastern and southwestern rims of the remnant \citep{koyama95}. 
Fig.~\ref{fig-sn1006} illustrates this with \xmm\ observations.
This strongly suggested that synchrotron radiation is the dominant
source of X-ray emission.
The spectral index of $\sim 3$ \citep{vink00a,allen01,dyer01}
is steeper than expected from
Fermi shock acceleration, indicating that the photons were
emitted by electrons with energies close to the maximum electron cosmic ray
energy.

The energy cut-off itself depends on the magnetic field strength
and is approximately, for a photon cut-off energy of 1~keV,
\begin{equation}
E_{\rm e} = 230/\sqrt{B_{\mu}}~{\rm TeV},
\label{eq-max-energy}
\end{equation} 
where $B_{\mu}$ is the magnetic field strength in $\mu$G.
As the interstellar magnetic field strength is $\sim 6~\mu$G, and a
shock compression factor of 4 may enhance it up to 20~$\mu$G,
the implied electron cosmic ray cut-off energy is around 50~TeV.
However, there is some dispute about the actual magnetic field strengths
in SNRs, as will be addressed in the next section.

After \snts\ X-ray synchrotron emission has been identified in a few
other supernova remnants, notably \casa\ \citep{allen97xx}, 
\rxj\ \citep{koyama97}, \rcwes\ \citep{borkowski01,bamba00}, 
and G266.2-1.2 \citep{slane01a}. 
I will return to some of those objects below.
Some of them have X-ray emission dominated by synchrotron radiation.
However, the X-ray emission from \casa\ is dominated by line emission,
but a non-thermal tail of emission is seen at hard X-rays,
which has been attributed to either synchrotron radiation or non-thermal
bremsstrahlung \citep[][]{allen97xx,favata97,bleeker01,vink03a}.

\begin{figure}[h]
	\centerline{
		\psfig{figure=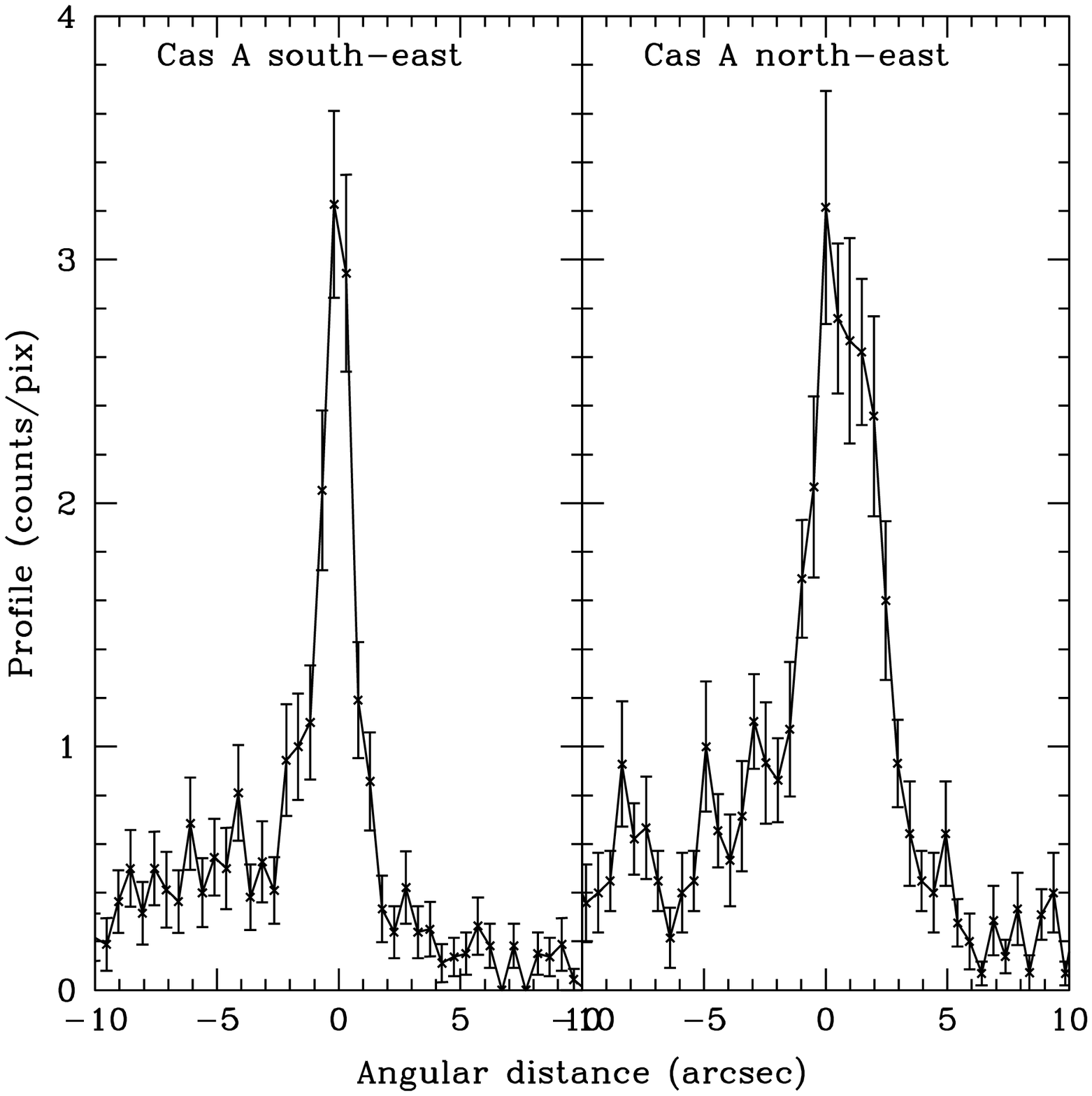,width=0.48\textwidth}\hskip 0.02\textwidth
		\psfig{figure=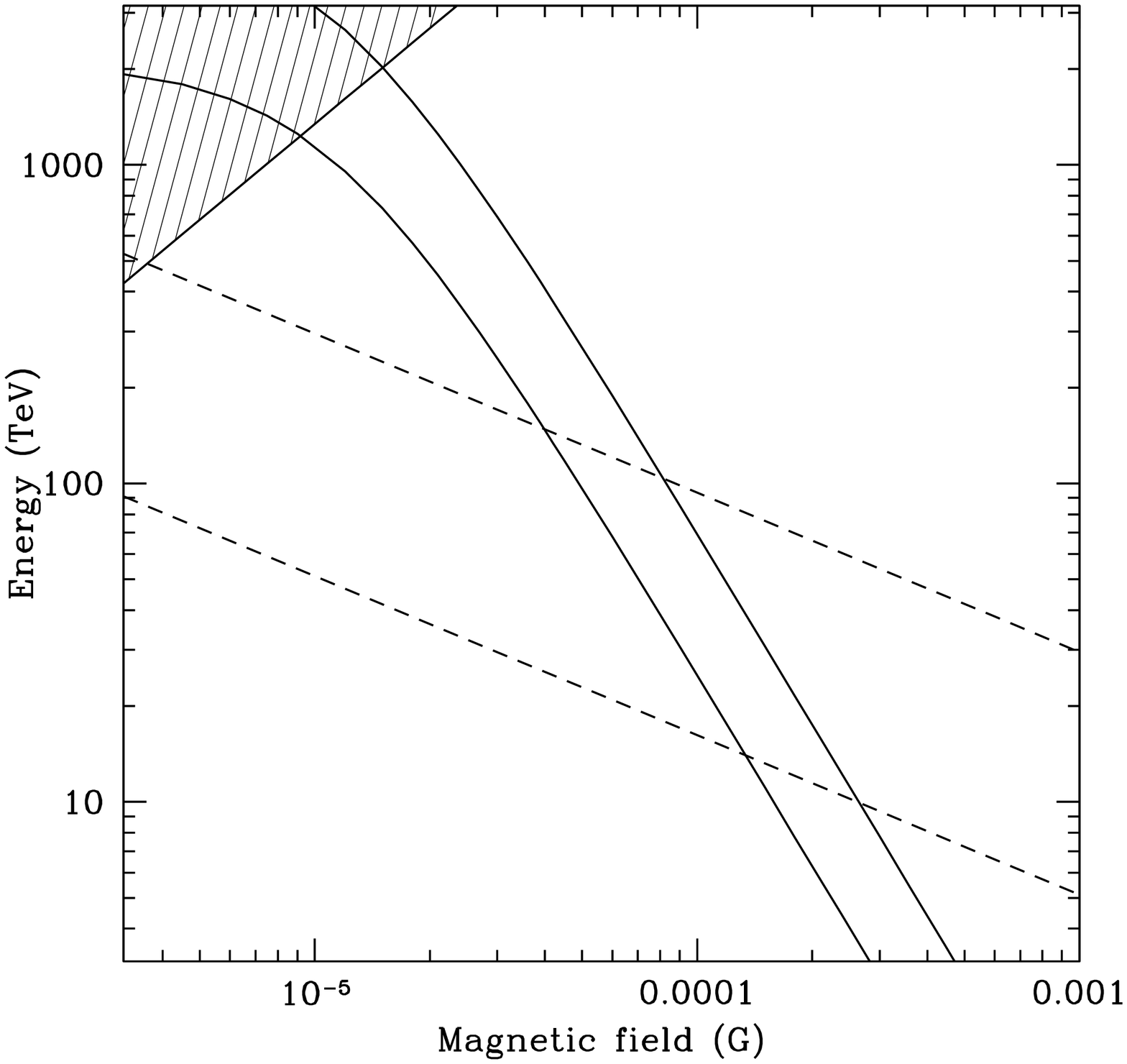,width=0.48\textwidth}
	}
\caption{On the left: 
	X-ray continuum emission profiles from two distinct
	regions at the outer rim of Cas~A (based on \chandra\ data). 
	On the right: the magnetic field strength and electron energies 
	derived from the width of the rims and the expansion velocity 
	\citep{vink03a}. See text for details.\label{fig-magn}
	} 
\end{figure}

\section{Cosmic ray acceleration and magnetic fields in SNRs}
The detection of synchrotron radiation in \snts\ and other remnants
has important consequences. On the one hand it is further proof
that SNRs accelerate electrons, as the high electron cut-off indicates
that the electrons can only have been accelerated recently.
On the other hand, $\sim 100$~TeV seems to be the maximum attainable electron
energy \citep{reynolds99}, 
which is one order of magnitude lower than the energy at
the ``knee'' of the cosmic ray spectrum.

The severity of the problem depends, however, on the nature of the
cut-off of the electron cosmic ray spectrum.
The ion and electron acceleration process is identical, apart from
the initial cosmic ray injection, which means that
if the energy cut-off is determined by 
the acceleration efficiency alone (i.e. the spectrum is age limited), 
the maximum electron energy is similar to the maximum ion energy.
The electron energy cut-off is much less of a problem, if the magnetic
fields in SNRs are greatly enhanced, and the spectra are loss limited, 
as losses affect predominantly the electron cosmic ray spectra.
The relative importance of loss limitations to age limitations depends
on the mean magnetic field strength of the medium through which the
electrons move.

An additional effect of higher magnetic fields is that the acceleration is more
efficient and age limitations of the ion cosmic ray spectrum can result in
maximum energies beyond the ``knee'' energy.
For that reason \citet{biermann93} proposed that most of the cosmic rays
around the ``knee'' are produced in SNRs moving through the stellar winds
of their progenitors, which may have enhanced the local magnetic field
with respect to the average galactic magnetic field.
An alternative idea is that the plasma waves generated by the cosmic rays
result in non-linear behavior, enhancing the background magnetic
field \citep{bell01}.
In other words, once cosmic rays are present they may
significantly enhance further cosmic ray acceleration.

\citet{vink03a} proposed a way of measuring the magnetic field
strength near the shock front of \casa. 
Independently \citet{bamba03} proposed the same method for \snts.
As an example I will concentrate on \casa.

The main idea is that the narrow rim of X-ray continuum emission surrounding
\casa\ is indeed X-ray synchrotron emission \citep{gotthelf01a}.
Near the shock front, electrons are continuously being accelerated, but as soon
as electrons move away from the shock front, radiative losses make that the
electron population cuts off at progressively lower energies, until it
does no longer produce X-ray emission.
The plasma moves downstream of the shock front with a velocity
of $\frac{1}{4}~v_s$. So the timescale, $\tau$, for radiative losses is coupled
to a length scale by $\Delta r = \frac{1}{4}v_s \tau$.

\begin{figure}
\parbox{0.5\textwidth}{
		\psfig{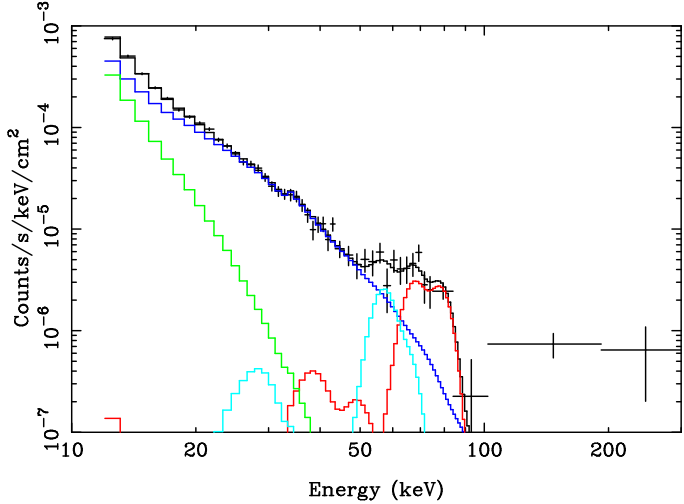}}
\hskip 0.05\textwidth
\parbox{0.45\textwidth}{
\vskip 4.5cm
\caption{\sax\ spectrum of \casa, 
	with a model consisting of combination
	of thermal and non-thermal bremsstrahlung, 
	$^{44}$Sc radio-active decay emission,
	and some intrumental line contributions 
	\citep{vink01a,vink03a}.\label{fig-casa-sax}}}
\end{figure}

The shock velocity of \casa\ has been measured to be $\sim 5000$~km/s 
\citep{vink98a,delaney03}, 
whereas the width of the rim, i.e. the length scale $\Delta r$, is between
1.5\arcsec\ to 4\arcsec.  
For a distance of \casa\ of 3.4~kpc \citep{reed95}, this corresponds
to loss times of 18~yr to 50~yr.
The loss times are inversely proportional to $B^2E_{\rm e}$, 
whereas the photon energy corresponding to the electron energy ($E_{\rm e}$)
scales as $B E_{\rm e}^2$.
Combining the measured loss times and the fact that X-ray synchrotron
radiation is observed around 5 keV allows the determination of the typical
magnetic field strength near the shock front, which is
about $10^{-4}$~G (Fig.~\ref{fig-magn}), more than an order of magnitude 
higher than the average galactic field strength.
Note that this approach assumes that the electrons at those high energies
are still coupled to the plasma, 
i.e. the mean free particle path, $\lambda_{mfp}$,
should be smaller than the rim width. 
In other words, the diffusion coefficient needs to be small.
As this is a prerequisite for acceleration at those energies, this seems
a reasonable assumption.
In order to be self-consistent the mean free path in a radial direction
should be larger than the gyroradius, $r_g$, for the derived magnetic.
The mean free path is often parameterized as $\lambda_{mfp} = \eta r_g$,
with $\eta \geq 1$ \citep[c.f.][]{reynolds98}. 
The width of the rim implies $\lambda_{mfp} < 10^{17}$~cm$^2$, or 
$\eta < 100$ for $B=0.1$~mG and $E\sim$50~TeV.

The relatively high magnetic field in \casa\ and \snts\ suggests 
that rapid cosmic ray acceleration of electrons and ions is 
possible,
and, as the electron population is loss limited, the ion cosmic rays can
in principle be accelerated up to, or beyond the ``knee''.
In fact, close to the shock front even electrons may obtain high energies, as
the electron energy indicated by  Fig.~\ref{fig-magn} applies to the
energy of the population downstream of the shock.

\citet{vink03a} also used another method for estimating the average magnetic 
field strength inside \casa, which consisted in comparing an upper limit on 
the non-thermal bremsstrahlung above 100~keV, assuming it is caused by the 
low energy part of the
electron cosmic ray spectrum, and the overall radio flux 
\cite[c.f.][]{cowsik80,atoyan00b}.

The bremsstrahlung normalization scales with 
$n_{\rm e} \Sigma_i n_{\rm i} Z_i^2$,
whereas the radio synchrotron emission scales with 
$n_{\rm e} B_{\perp}^{(\Gamma+1)/2}$.
As $\Sigma_i n_{\rm i} Z_i^2$ is approximately known, 
the average relativistic electron density and magnetic field strength can be inferred.
The measurements imply $B > 0.5$~mG,
higher than the magnetic field strength near the shock.
The reason is presumably magnetic field enhancements due to turbulence 
associated with the contact discontinuity of shocked ejecta and
swept up circumstellar matter. This may explain why the brightest radio
structure in \casa\ is a shell coinciding with the ejecta shell, 
instead of with the outer rim.

The high overall magnetic field makes it unlikely that X-ray synchrotron
radiation is coming from the bright radio and X-ray shell of \casa; 
inside the bright shell the synchrotron loss time for the X-ray 
emitting electrons is only of the order of 10~yr, much less than the age
of \casa.
As some hard X-ray  emission seems to be associated with that bright
shell \citep{bleeker01},
it is likely that another emission mechanism is contributing to the
hard X-ray emission. 
\cite{laming01a,laming01b} has proposed that internal shocks 
produce lower hybrid plasma waves, which are responsible
for accelerating electrons up to $\sim 100$~keV, 
where the observed X-ray spectrum has a cut-off.
This predicted spectrum was found to be consistent with the hard X-ray spectrum
obtained with a deep, 500~ks,  \sax\ observation of \casa\ 
\citep[][Fig.~\ref{fig-casa-sax}]{vink03a}.

\section{TeV emission: Evidence for hadronic cosmic ray acceleration or not?}

So far our discussion has focused on electron cosmic rays, 
as they are most easily observed through synchrotron radiation and 
bremsstrahlung.
However, hadronic cosmic rays are arguably the most important part of the
cosmic rays, as they constitute the cosmic rays best observed at earth, and 
the ratio of hadronic to leptonic (electron) cosmic rays 
is roughly $100 : 1$.
The direct observation of ultra-relativistic in SNRs is 
therefore an important part of the proof that shock acceleration in 
SNRs is responsible for the observed cosmic ray spectrum up to the ``knee''.
This goal is coming nearer, or may even have been reached, with the recent
advances made with telescopes observing Cherenkov radiation in the earth
atmosphere caused by TeV \gray s.

Although the electromagnetic radiation from relativistic ions is weak,
ions colliding with background ions betray themselves by producing pions 
($\pi^+, \pi^-, \pi^0$), which decay quickly into muons, positrons, 
electrons and
neutrinos, except for $\pi^0$, which decays in 99\% of the cases 
into two photons with energies of 68~GeV each in the $\pi^0$\ rest frame.
The threshold for pion creation is 290~MeV. 
Well above the production threshold 30\% of the particle energy will be 
transferred to the pion, and the emerging \gray\ spectrum will have a 
power law slope similar to
the index of the particle spectrum.

Pion decay is the dominant component of the galactic background \gray\ emission
above 1~GeV \citep[][]{hunter97}. 
However, the situation for \gray\ emission from SNRs is not so clear.
First of all, \egret\ detected only few of the  known SNRs, most of which 
turn out to be older SNRs, like IC 443 and $\gamma$-Cygni \citep[see][for a 
review]{Torres03}.
Secondly, inverse Compton scattering of background photons by ultra-relativistic
electrons is a very plausible, alternative, explanation
for TeV \gray\  emission from young SNRs.

\begin{figure}
\parbox{0.5\textwidth}{
\psfig{figure=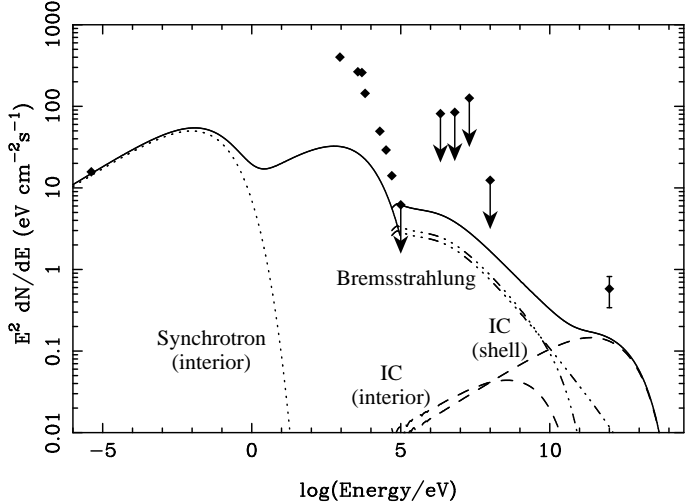,angle=-90,width=0.5\textwidth}}
\hskip 0.05\textwidth
\parbox{0.45\textwidth}{
\caption{
	Observed continuum emission from \casa (data points) and
	a simple model showing the expected contributions from
	electron cosmic rays (solid line).
	The emission is separated in two zones; 
	1) from a region near the outer shock  with $B \sim 0.1$~mG, 
	which dominates the emission above 100~keV, and 
	2) from the interior of the remnant ($B > 0,5$~mG),
	which dominates	the radio synchrotron emission.
	The dashed line indicates inverse compton (IC), 
	and the dashed-dotted line the bremsstrahlung
	contributions. The contributions from the shell
        and interior are separated.
	\citep[Adapted from][]{vink03a}.\label{fig-casa-gamma}}}
\end{figure}

The first shell-type SNR detected above energies of $1.7$~TeV is
\snts\ \citep{tanimori98}. 
The discovery was made with the \cangeroo\ telescope.
Less than two years before this detection the synchrotron nature of the X-ray
emission had been established, so it was natural to suggest that
the same electrons responsible for the X-ray synchrotron radiation
were responsible for the \gray\ emission by means of 
inverse Compton scattering of cosmic microwave background (CMB) photons.
As the normalization for inverse Compton scattering 
scales with $n_e n_{ph,CMB}$, 
and synchrotron radiation with  $n_e B^{(\Gamma+1)/2}$,
with $n_{ph,CMB}$ the photon density, one can in principle
infer the magnetic field from the ratio of the hard X-ray and TeV \gray\
emission,
which turns out to be $B = 6.5\pm2~\mu$G \citep{tanimori98}.

However, this is only valid if the TeV emission is indeed due 
to inverse Compton scattering.
\cite{berezhko02} recently argued that from a theoretical point of view 
a higher magnetic field is preferred, 
a point of view that seems to be supported by the
magnetic field measurement by \citet{bamba03}.
This would mean that the relativistic electron density necessary to
explain the X-ray synchrotron spectrum can be lower,
which decreases the inferred inverse Compton scattering contribution
and the maximum electron energy (Eq.~\ref{eq-max-energy}).
\cite{berezhko02} argue that therefore pion decay is the likely
origin of the TeV \gray\ emission from \snts.
An additional argument against dominant inverse Compton scattering above 1~TeV
is that the  \gray\ emission seems to come 
from the northeastern side of \snts, whereas
the synchrotron radiation, 
caused by the same electrons that upscatter background
photos, is observed from both the northeastern and southwestern rims
of the remnants (Fig.~\ref{fig-sn1006}).
In the case of pion decay this asymmetry could be explained by the presence of
a density enhancement in the vicinity of the northeastern rim.

Two other young remnants have been discovered by Cherenkov telescopes.
\casa\ was detected at the 4.9$\sigma$\ level with the \hegra\ telescope
\citep{aharonian01}.
For \casa\ the case for pion decay as the origin for the TeV emission
is even more compelling than for \snts, although still not conclusive.
As explained in the previous section, the average magnetic field
inside \casa\ is $>0.5$~mG, which means that the electron synchrotron
loss time is short, making it unlikely that most electron populations
inside \casa\ extend all the way up to 100~TeV, even if most
of the hard X-ray emission is synchrotron radiation.
This makes inverse Compton emission a less likely mechanism for producing
TeV emission.
However, as \citet{vink03a} have shown (see previous section), 
the magnetic field near the shock front is $\sim 0.1$~mG, and the relativistic
electron density is somewhat higher.
Moreover, \casa\ itself is a bright infrared source, so that more background
photons are present, 
which have on average a higher energy as well \citep{atoyan00b}.
This gives rise to the additional complication of an anisotropic photon field.
Despite these reservations, \cite{vink03a} argue that pion decay and not
inverse Compton emission is the likely source for the TeV emission, 
but not by a wide enough margin to be completely confident that the inverse 
Compton scattering interpretation
can be excluded (Fig.~\ref{fig-casa-gamma}).

Of the three remnants detected above 1~TeV, \rxj\ is the most surprising
one. The remnant was discovered with \rosat\ \citep{pfeffermann96},
and \asca\ spectra showed that the X-ray emission 
was dominated by non-thermal radiation \citep{koyama97}.
The detection of TeV emission was made with the \cangeroo\ telescope 
\citep{muraishi00}.

One of the interesting features of \rxj\ is that it 
is probably an old remnant,
judged by its size of $\sim 34$\arcmin\ and the likely
association with a molecular cloud at a distance of roughly 6~kpc
\citep{slane99}, whereas it is still a source of X-ray synchrotron radiation. 
The cosmic ray spectra in older remnants is likely to be loss limited,
but the maximum {\em electron} energy is proportional to the shock
velocity for loss limited models \citep{reynolds98}, which means
that older remnants are not expected to produce
considerable X-ray synchrotron radiation.

As pointed out by \citet{slane99} some of this problem may be
alleviated, if \rxj\ is a remnant evolving in stellar wind cavity.
In that case the shock velocity remains high until
it encounters the cavity wall, after which the remnant is believed to
brighten rapidly and radiate most of its energy away in a relatively
short time.
Another surprising feature of this remnant is that no evidence has
turned up yet that there is thermal X-ray emission coming from this remnant,
unlike \snts.
This is only possible, if either the electron temperature is very cool
(below 0.1~keV), or the density inside the remnant is very low,
(cf. \snts\ which has $n_{\rm e} \sim 0.1$~cm$^{-3}$). 

The fact that the X-ray emission from \rxj\ is synchrotron dominated
suggested, as it did for \snts, that the emission above 1~TeV is caused
by inverse Compton emission.
However, recent observations with \cangeroo\ showed that the 1-10~TeV \gray\
spectrum, in  combination with the X-ray spectrum, is incompatible
with inverse Compton emission, and that it is therefore likely that
the \gray\ emission is the result of pion decay \citep{enomoto02}.
The \cangeroo\ \gray\ spectrum would therefore be the first direct 
evidence that nucleons are accelerated by SNRs.
However, shortly after the announcement that ultra-relativistic ions
had finally been observed in SNRs 
both \citet{reimer02} and \citet{butt02} pointed out
that the spectrum of a nearby and possibly associated
\egret\ source, 3EG~J1714-3857, was not consistent with a pion decay
spectrum from \rxj, as the \egret\ data points lie below the pion decay
model.
Although an obvious objection is that 3EG~J1714-3857 and \rxj\
do not spatially coincide, it is hard to believe that \egret\ detected
3EG~J1714-3857, whereas a brighter nearby source should have been missed.

\section{The cosmic ray injection efficiency}
One topic not yet addressed in this article is the injection process and injection efficiency
of cosmic rays. Our knowledge of the injection process, especially for electrons, 
is limited, but some progress has been made over the last decade 
\citep[e.g.][]{bykov99}, and computers are now powerful enough 
to simulate the microscopic processes with particle in cell methods \citep{schmitz02b,schmitz02a}.

The cosmic ray efficiency is also of major importance for the hydrodynamical modeling of
supernova remnants, because, if a major fraction of the shock energy goes into accelerating
particles, the plasma may become cosmic ray dominated, possibly increasing the shock compression
ratio and lowering the mean gas temperature (Eq.~\ref{eq-shocks}).
If the acceleration is very efficient, high energy particles may diffuse away from the remnant
altogether, making the shock essentially radiative, under which conditions the compression
ratio may even exceed a factor of 7 \citep[e.g.][]{decourchelle00}.
This has in fact been claimed for the Small Magellanic Cloud remnant 1E~0102.2-7219,
for which \cite{hughes00b} has shown that the electron temperature is lower than can be
expected, based on the observed shock velocity, even when allowing for slow electron-ion temperature
equilibration.

However, it should be pointed out that in most cases, even if very efficient
cosmic ray acceleration takes place, it is not likely that the shocks becomes radiative
or even acts like a relativistic gas with $\gamma=4/3$. 
The reason is that most cosmic ray spectral indexes are observed to be $\Gamma >2$, for which 
the total cosmic ray energy\footnote{This follows from the integration over the cosmic ray spectrum, 
$\int_{E_0}^{\infty} \kappa E^{-\Gamma}EdE$.} is dominated by non-relativistic cosmic rays.
This can be illustrated for \casa, as the magnetic field strength measurement
allows for the estimation of the electron cosmic ray density normalization, which turns out to be
$\kappa \sim 5\times10^{-11}$\ (cgs).\footnote{See the previous footnote. $\kappa$\ normalizes the
total number of electron cosmic rays per cm$^{-3}$ (for \casa\ $\Gamma = 2.55$).}
If we integrate the number density, assuming a power law spectrum in momentum,
we find that the number of particles above $p_0 = 0.01 m_{\rm p}c$\ 
\citep[somewhat arbitrary, but following][]{bell87} is
$\sim 8\times10^{-4}(\zeta+1)$, with $\zeta$\ the ratio of nucleonic to 
electron cosmic rays.
As the thermal particle density in \casa\ is about 20~cm$^{-3}$\ and $\zeta$
is generally believed to be in the range 1-100, we find a cosmic ray injection
efficiency between $\phi = 4\times10^{-5}$\ to $4\times10^{-3}$. 
Interestingly, this range in injection efficiencies
coincides with the transition from thermally dominated pressure to
cosmic ray dominated pressure \citep{bell87}, 
suggesting that perhaps the acceleration process is 
indeed self-regulating. Note, however, 
that for the steep cosmic ray spectrum of \casa\ the cosmic ray pressure is 
dominated by non-relativistic nucleonic cosmic rays. 

Our understanding would be greatly advanced if we could observe the injection 
spectrum directly.
\cite{vink97} suggested that the the X-ray spectrum from certain parts of 
\rcwes, which
are characterized by featureless spectra except for the presence of iron line 
emission at 6.4~keV,
could be explained by a low temperature electron gas with a non-thermal 
bremsstrahlung tail, 
which also causes line emission from underionized iron 
\citep[see also][]{vink02b}.
This non-thermal tail may be caused by the electron cosmic ray injection 
spectrum. However, \citet{rho02} have shown that such a bi-modal electron 
distribution will quickly evolve toward a Maxwellian distribution, 
resulting in more line emission. 
They prefer  to explain the
observed spectral features by assuming that the continuum is X-ray synchrotron radiation,
whereas part of the plasma inside the remnants consists of a hot, low density, pure iron gas 
\citep[see also][]{bamba00,borkowski01,bykov02}.

\section{Concluding remarks}
I have reviewed the current observational evidence for efficient cosmic ray acceleration
by the collisionless shocks in supernova remnants.
For a long time the evidence consisted only of the observed non-thermal radio spectra, but the
last decade the more important evidence has come
from X-ray and \gray\ observations, which indicate that electrons are accelerated
up to 100~TeV.
However, direct evidence that SNRs accelerate cosmic rays  up to, or beyond, the ``knee'',
and that both electrons and ions are accelerated,
is not unambiguous yet. 
This situation
is likely to improve considerably over the next five years, 
with new space missions like \integral\ and \glast, 
and the completion of new Cherenkov telescopes, such as \hess.

\section*{Acknowledgements}
It wish to express my gratitude to Martin Laming, Johan Bleeker, and Jelle Kaastra, 
my collaborators on some of the topics discussed here.
This work is supported by the NASA
through Chandra Postdoctoral Fellowship Award Number PF0-10011
issued by the Chandra X-ray Observatory Center, which is operated by the
Smithsonian Astrophysical Observatory for and on behalf of NASA under contract
NAS8-39073.

\renewcommand{\refname}{References}
\setlength{\bibspacing}{0.1\baselineskip}

\vskip 0.5cm
\noindent
\end{document}